\documentclass[12pt,preprint]{aastex}

\newcommand{\etal}{et~al.}
\newcommand{\fig}{Fig.~}
\newcommand{\mic}{\,$\mu$m}
\newcommand{\av}{A_V}
\newcommand{\rv}{R_V}
\newcommand{\ejk}{E_{J-K}}
\newcommand{\ejh}{E_{J-H}}
\newcommand{\ehk}{E_{H-K}}
\newcommand{\kmag}{K_{\rm s}}
\newcommand{\water}{H$_2$O}

\slugcomment{Version date: \today}

\begin{document}

\title{A CATALOG OF BACKGROUND STARS REDDENED \\ BY DUST IN THE TAURUS DARK CLOUDS \\ ~}
\author{S.~S. Shenoy$^{1,2}$, D.~C.~B. Whittet$^1$, J.~A. Ives$^1$, and D.~M. Watson$^3$}

\altaffiltext{1}{Department of Physics, Applied Physics \& Astronomy,
	Rensselaer Polytechnic Institute, 110 Eighth Street, Troy, NY 12180.}
\altaffiltext{2}{Present address: Spitzer Science Center, Mail Code 220--6,
California Institute of Technology, Pasadena, CA 91125. Email: sshenoy@ipac.caltech.edu.}
\altaffiltext{3}{Department of Computer Science,
	Rensselaer Polytechnic Institute, 110 Eighth Street, Troy, NY 12180.}

\begin{abstract}

Normal field stars located behind dense clouds are a valuable resource in interstellar astrophysics, as they provide continua in which to study phenomena such as gas-phase and solid-state absorption features, interstellar extinction and polarization. This paper reports the results of a search for highly reddened stars behind the Taurus Dark Cloud complex. We use the Two Micron All Sky Survey (2MASS) Point Source Catalog to survey a $\sim$\,50~deg$^2$ area of the cloud to a limiting magnitude of $K_{\rm s} = 10.0$. Photometry in the 1.2--2.2\mic\ passbands from 2MASS is combined with photometry at longer infrared wavelengths (3.6--12\mic) from the {\it Spitzer Space Telescope\/} and the {\it Infrared Astronomical Satellite\/} to provide effective discrimination between reddened field stars and young stellar objects (YSOs) embedded in the cloud. Our final catalog contains 248~confirmed or probable background field stars, together with estimates of their total visual extinctions, which span the range $2<\av<29$~mag. We also identify the 2MASS source J04292083+2742074 (IRAS~04262+2735) as a previously unrecognized candidate YSO, based on the presence of infrared emission greatly in excess of that predicted for a normal reddened photosphere at wavelengths $>5$\mic.

\end{abstract}


%\clearpage
\section{Introduction}
\label{intro}

Observations of many interstellar phenomena rely on the presence of background field stars, i.e.\ stars lying beyond the region of interest that provide sources of continuum radiation. Examples include absorption-line spectroscopy of ionic, atomic and molecular gas, spectroscopy of solid-state absorption features in dust grains, and the continuum effects of interstellar extinction and polarization introduced by dust, at wavelengths extending from the far ultraviolet to the mid-infrared. Observations at ultraviolet and visible wavelengths are generally limited to lines of sight with low to moderate extinction and therefore sample mostly diffuse phases of the interstellar medium (ISM). Observations in the near to mid-infrared allow studies of these absorptive phenomena to be extended to dense molecular clouds, where they provide information complementary to that obtained at much longer wavelengths were cool dust and gas can be observed in emission in their own right. 

Observing absorption features in discrete lines of sight toward background stars places an obvious limitation on spatial resolution. Consider, for example, the distribution of molecular gas in cold, dense clouds, and the depletion of molecules onto the surfaces of dust grains to form icy mantles. Radio astronomers can readily map the distribution of molecular gas by observing its intensity in a chosen spectral emission line (most commonly of CO) at a spatial resolution limited only by the beam of the telescope (see Mizuno \etal\ 1995 for an example relevant to the current work). However, to map the corresponding distribution of molecular solids one must study an infrared absorption feature, such as the 3.0\mic\ feature of solid H$_2$O, against a continuum provided by a background star. The first attempt to map the ice distribution by this method, that of Murakawa \etal\ (2000), was limited to detections of the 3.0\mic\ feature in some~25 lines of sight in the Heiles Cloud~2 region of Taurus. At mid-infrared wavelengths, it is possible to study interstellar absorptions against extended background emission, as described by Sonnentrucker \etal\ (2008) for the 15\mic\ feature of CO$_2$; but this opportunity arises only in a limited number of cases, where cold material is located fortuitously in front of warmer material heated to appropriate temperatures by nearby luminous stars. In regions of low-mass star formation a point-like stellar continuum source is generally the only option. Interstellar absorption features are observed routinely in the spectra of young stellar objects (YSOs) embedded in the clouds (e.g.\ Boogert \etal\ 2004), but in these lines of sight ambiguity often exists between interstellar matter and material local to the source that may be modified by its presence. Only background stars provide the means of studying absorptions arising in quiescent regions of the dense ISM, remote from sources of radiation. 

A primary motivation for infrared surveys of dark clouds is to conduct a census of star formation, in which YSOs are carefully distinguished from field stars. The Taurus dark-cloud complex is a key region in studies of low-mass star formation (e.g.\ Luhman \etal\ 2006; G\"udel \etal\ 2007; Padgett \etal\ 2007), and also a valuable laboratory for studying interstellar molecules and dust (e.g.\ Pratap \etal\ 1997; Dickens \etal\ 2001; Whittet \etal\ 2001). It is nearby ($\sim 140$\,pc; Kenyon \etal\ 1994; Loinard \etal\ 2005) and situated at moderate Galactic latitude ($b \sim 15^\circ$), circumstances that aid both types of study: there is relatively little risk of confusion between stellar associations at different distances along the line of sight, compared with a region in the Galactic disk, and almost all of the interstellar extinction and absorption in the direction of the cloud arises in the cloud itself. The first infrared survey of the Taurus region to include a significant number of reddened background stars was that of Elias (1978), who identified about a dozen such objects that have since been observed intensively for interstellar absorption features (e.g.\ Whittet \etal\ 2007 and references therein). 

In recent years, the task of identifying reddened field stars has been greatly facilitated by the availability of data from the {\it Two-Micron All-Sky Survey\/} (2MASS; Skrutskie \etal\ 2006) and from surveys at longer infrared wavelengths made with the {\it Spitzer Space Telescope\/} (e.g.\ Luhman \etal\ 2006). The goal of this paper is to use these resources to construct a much larger catalog of field stars reddened by dust in the Taurus dark-cloud complex than has previously been available.

\section{Photometric data and associations}
\label{dataset}
\subsection{2MASS}
The survey area selected for this work, shown in \fig 1, is based on the \(^{13}\)CO (J=1--0) map of the Taurus cloud complex reported by Mizuno \etal\ (1995, their \fig 3). Near-infrared photometric data in the 1.25\mic\ ($J$), 1.65\mic\ ($H$) and 2.17\mic\ ($\kmag$) passbands were compiled for stars within this survey area from the 2MASS survey\footnote{The data were collected using the Gator web interface of the Infrared Science Archive operated by the Infrared Processing and Analysis Center (http://irsa.ipac.caltech.edu/applications/Gator/).}. A limiting magnitude of $\kmag = 10.0$ was adopted: this constraint ensures that only stars with highest-quality photometry are selected, and has the additional benefit of excluding background red dwarfs, thus avoiding a possible source of ambiguity in the intrinsic colors (see \S4). Further quality control was provided by the various flags included in the 2MASS catalog. Only observations with read flag (rd\_flg) values of 1, 2 or 3 were selected as being high quality detections with reliable astrometry and photometry. Other flags that deal specifically with the quality of the data are the photometric quality flag (ph\_qual) and the contamination and confusion flag (cc\_flg). We selected only data with a ph\_qual value of ``A" in each passband, signifying the best quality photometric data one can obtain with 2MASS. The cc\_flag indicates whether the photometry and/or positional measurement of a source may be contaminated or biased due to the proximity to an image artifact or a nearby source of equal or greater brightness: we chose only objects with cc\_flg~$= 0$, indicating sources unaffected by known artifacts or confusion. Photometric errors in the resulting data are $\pm 0.03$~mag or less in each passband.

The following color constraints were used to discriminate against unreddened stars and stars with anomalous colors in the 2MASS database:
\begin{equation}\label{rvl}(J-H) > 1.75\,(H-\kmag) - 0.04\end{equation} 
\begin{equation}\label{rvu}(J-H) < 1.75\,(H-\kmag) + 0.56\end{equation}
\begin{equation}\label{hkc}H-\kmag > 0.4.\end{equation}
The multiplicative constant 1.75 in eqs.~\ref{rvl} and \ref{rvu} is based on the observed reddening law (see \S3): these equations define the range of colors over which normal stars may be dereddened onto intrinsic color lines (Itoh \etal\ 1996), as illustrated in \fig 2a. Discrimination against normal stars with little or no reddening is provided by Eq.~\ref{hkc}, which effectively excludes those with visual extinctions $\av<3$~mag. This extinction limit corresponds approximately to the threshold extinction for detection of ices in the region (Whittet \etal\ 2001 and references therein).

Application of the above criteria resulted in selection of 293~sources. Stellar associations were sought in the SIMBAD database, and it was found that the sample included a substantial number of sources associated with previously known YSOs. As an additional check on the status of each target, our list was collated with comprehensive catalogs of YSOs in the Taurus region recently compiled from infrared and X-ray observations (Luhman \etal\ 2006; G\"udel \etal\ 2007; Scelsi \etal\ 2007). A total of 40~known or probable YSOs and 4~variable stars of other types were excluded, reducing the sample to 249~sources. This final catalog is presented in Table~1, and the distribution of the sources on the sky is shown in \fig 1. 2MASS identifications and photometry are listed in Table~1, together with IRAS associations, stellar associations and spectral types, as available from the literature, and addition infrared data described below.

\subsection{IRAC}
Images from the Infrared Array Camera (IRAC; Fazio \etal\ 2004) on board the {\it Spitzer Space Telescope\/} provide photometry in four passbands centered at 3.6, 4.5, 5.8 and 8.0\mic. The star-formation survey reported by Luhman \etal\ (2006) is based on a photometric catalog containing some 450,000 sources in the Taurus region (the Taurus IRAC Point Source Archive), of which photometry for only $\sim 160$~confirmed YSOs has been published to date. The IRAC data were processed by the Wisconsin IRAC Pipeline developed for the GLIMPSE Galactic Plane Survey and the SAGE Large Magellanic Cloud Survey (see Luhman \etal\ 2006 for further details). The entire catalog was kindly made available to us by Barbara Whitney and Marilyn Meade, and is used here in combination with the 2MASS data to provide an additional constraint on the nature of our sample of candidate field stars (\S3). 

The IRAC Point Source Archive was supplied in ascii format and a custom web-based interface was constructed to facilitate access. Collation with our field star catalog yielded associations for 186 out of 249 objects: the area of sky covered by the IRAC observations (\fig 1 in Luhman \etal\ 2006) is somewhat smaller than our survey area, thus some of our 2MASS candidates lack IRAC coverage. All available photometry is included in Table~1. A few of the brightest sources are saturated at 3.6 and 4.5\mic. Photometric errors are typically $\pm 0.05$~mag or less in each passband.

\section{Color-color diagrams}
\label{colors}
The $J-H$ vs.\ $H-K$ color-color diagram provides useful but imperfect discrimination between dust-embedded stars and stars with normal photospheres subject to reddening (see Itoh \etal\ 1996 and Gutermuth \etal\ 2004 for discussion and examples). $J-H$ is most sensitive to the photospheric temperature of the star, whereas $H-K$ is also sensitive to emission from a circumstellar shell or disk~--- provided the circumstellar material is sufficiently warm. A more stringent discriminant that recognizes the presence of cooler circumstellar matter is possible in cases where additional photometry at longer wavelengths is available: we follow Gutermuth \etal\ (2004) in adopting the $J-H$, $H-[4.5]$ diagram as a valuable complement to $J-H$, $H-K$. Both diagrams are plotted for our candidate field stars in \fig 2. As a control sample, we plot confirmed Taurus YSOs on the same axes in \fig 3, using data from Luhman \etal\ (2006). Photometric errors are typically comparable with or smaller than the size of the plotting symbol in all figures.

Diagonal lines in Figs.~2 and 3 are parallel to the expected displacement caused by interstellar reddening, and delimit the approximate area occupied by normal stars: stars in this zone may be dereddened onto intrinsic color lines. In the case of $J-H$, $H-\kmag$, these lines (eqs.~1 and 2) have slope set to the reddening ratio $\ejh/\ehk = 1.75$ determined from observations of Taurus field stars (Whittet \etal\ 2007); the corresponding lines in $J-H$, $H-[4.5]$ assume the standard average interstellar extinction law (Whittet 2003) to transform from $H-\kmag$ to $H-[4.5]$. Intrinsic $J-H$ and $H-\kmag$ colors are from Bessell \& Brett (1988) with transformations from Carpenter (2001). Intrinsic $H-[4.5]$ colors are based on an interpolation between ground-based $H-L$ and $H-M$, colors and should be treated as no more than a rough guide to the loci of normal unreddened stars.

All of our candidate field stars fall within the ``normal reddened photospheres" zone of the $J-H$, $H-\kmag$ diagram (\fig 2a) by definition, as this was included in the selection criteria (\S2.1). \fig 2b shows that all with IRAC data available also lie within the corresponding zone of the $J-H$, $H-[4.5]$ plot. This is in contrast to the distribution of YSOs (\fig 3), many of which show a displacement toward the right, especially in $H-[4.5]$, characteristic of circumstellar excess emission affecting the longer-wavelength color index. We consider the distribution of our field-star candidates in \fig 2b to be strong evidence that the large majority are, indeed, normal reddened field stars. Perusal of the data in Table~1 indicates that this conclusion is independent of our choice of [4.5] as the representative IRAC passband: for example, the [4.5] and [8.0] values generally agree to within $\sim 0.2$~mag, indicating no gross differences between $H-[4.5]$ and $H-[8.0]$.\footnote{J04292083+2742074 is the only major exception (see \S5).}

As a further test, we plot in \fig 4 the $J-H$ vs.\ $H-[12]$ color-color diagram for 26~stars that have 12\mic\ data available (see column~11 of Table~1 and associated footnotes). Photometric values were calculated from 12\mic\ fluxes taken from the IRAS Point Source Catalog, as available (19~sources), using the method described in the IRAS Explanatory Supplement (1988). Fluxes at 12\mic\ for seven additional stars were estimated from published spectra obtained with the Infrared Spectrometer (IRS) of the Spitzer Space Telescope (Whittet \etal\ 2007). YSOs catalogued by Luhman \etal\ (2006) that have IRAS associations are also included in \fig 4 for comparison. The zones occupied by YSOs and by reddened field stars are particularly well separated in this diagram, confirming the field-star status of all but one of the 26~candidates in the subset. A clear anomaly is identified in the case of J04292083+2742074 (IRAS~04262+2735), however: it plots with the field stars in \fig 2 and with the YSOs in \fig 4. Further discussion of this object is deferred to \S5.

The level of certainty with which field-star status is assigned to each object in Table~1 naturally varies according to the information available. A total of 188~stars have photometry in IRAC and/or 12\mic\ passbands consistent with an absence of circumstellar dust. Considering infrared data alone, this set might be confused with evolved (class~III) YSOs that have completed dispersal of their dusty envelopes; however, such stars are typically strong X-ray sources and these have been excluded from our catalog (\S2). Least securely characterized are 56~stars that lack spectral classifications and have photometry only in the 2MASS passbands.

\section{Extinction}
\label{ext}
The visual extinction ($\av$) was estimated for each star in Table~1 from 2MASS photometry. $\av$ is related to the infrared color excess $\ejk$ by the relation 
\begin{equation}
\av = r\ejk
\end{equation}
where $r$ is a factor that depends on the form of the extinction curve over the relevant wavelengths: a mean value $r=5.3\pm 0.3$ was determined by Whittet \etal\ (2001) for the Taurus cloud (for comparison, $r \approx 6.0$ in the diffuse ISM). For stars with known spectral classifications, $\ejk$ is determined routinely from observed and intrinsic colors to yield $\av$. For stars lacking spectral classifications, $\ejk = \ejh+\ehk$ is estimated from the observed locus in the $J-H$ vs.\ $H-\kmag$ diagram (\fig 2a) by extrapolation along the appropriate reddening vector onto intrinsic color lines. This generally provides unambiguous results, notwithstanding the separation of giant and dwarf intrinsic colors at late spectral types (\fig 2a). Background stars bright enough to be included in our sample are expected to be predominantly either late-type (K, M) giants, which deredden onto the upper branch, or main-sequence stars earlier than K0; red dwarfs distant enough to be background to the cloud are predicted to be too dim at 2.2\mic\ to be selected\footnote{Dwarfs with extinction $\av=3$~mag and spectral types K5, M0 and M5 are predicted to have $\kmag = 10.5$, 11.1 and 12.2~mag, respectively, at the distance of the cloud (140\,pc), and are thus excluded by our $\kmag=10$~mag limit (\S2).}. Note that our method for evaluating $\av$ improves upon a ``one fits all" intrinsic $J-K$ color adopted for background stars lacking spectral classifications in some previous literature (Tamura \etal\ 1987; Goodman \etal\ 1992). Several Taurus YSOs also have near infrared colors consistent with normal reddened photospheres (compare Figs.~2a and 3a): extinction estimates have been obtained for the 10~most reddened (those with $J-H > 2.5$), and results are listed in Table~2. Our $\av$ estimates for both field stars and YSOs are thought to be accurate to $\pm 0.5$~mag or better. 

A histogram of $\av$ values from Table~1 is plotted in \fig 5. The sample of 249~stars is divided into one--magnitude bins (black columns). Also shown is the effect of adding 81~optically-selected reddened stars ($\av>0.5$) in Taurus from the extinction studies of Straizys \& Meistas (1980) and Whittet \etal\ (2001), which cover a similar area of sky. The minimum occurring in the $\av=2$--3~mag bin is probably not real but a result of incomplete sampling: our color criteria for infrared-selected sources (\S2) introduced a sharp cutoff for $\av<3$, and the optically-selected sample is expected to become increasingly incomplete for $\av>2$. Overall, the distribution shows a broad peak centered near $\av=3.5$, with a tail extending to $\av\sim 10$, and sporadic higher values (19~field stars with $\av>10$). 

The distribution on the sky of stars from Tables~1 and 2 is shown in \fig 1, with extinction distinguished by plotting symbol in three groups: low ($2<\av<5$), intermediate ($5<\av<10$) and high ($\av>10$). As expected, lines of sight with high extinction generally cluster toward known condensations such as L1495, B18 and TMC-1. Lines of sight with low and intermediate extinction lie predominantly toward the outer boundaries of condensations, but some are more widely distributed. See Cambr\'esy (1999) and Padoan \etal\ (2002) for more detailed maps of extinction in the Taurus region based on optical star counts and statistical analysis of 2MASS data, respectively.

\section{J04292083+2742074: A candidate YSO}
\label{iras}
The 2MASS source J04292083+2742074 is identified as a possible YSO from its anomalous position in the $J-H$ vs.\ $H-[12]$ diagram (\fig 4), based on an association with IRAS point source 04262+2735. Perusal of the 2MASS database shows no other 2.2\mic\ source  within the error ellipse of the IRAS position of sufficient brightness to cause confusion, thus the association appears to be secure. To elucidate the nature of this object, its spectral energy distribution (SED) was constructed from all available photometry (2MASS, IRAC, IRAS) and plotted in \fig 6. The SED of the prototypical reddened field star Elias~16 is also shown for comparison. A fit to each SED was calculated assuming model atmospheres from Kurucz (1992) with extinction from the Cardelli, Clayton \& Mathis (1989) $\rv$-dependent empirical law ($\rv=\av/E_{B-V}$ is the ratio of total to selective visual extinction). For Elias~16, we assume Kurucz model t4500g20p00 (appropriate to a K giant with $T_{\rm eff} = 4500$~K) reddened to $\av=24$~mag with the $\rv=4.0$ extinction law. For IRAS~04262+2735, the same effective temperature was assumed and SEDs were calculated for both giant and dwarf spectra (Kurucz models t4500g20p00 and t4500g50p00, respectively). In practice is was found that the two models gave closely similar results: only that based on the dwarf intrinsic spectrum is shown in \fig 6, reddened to $\av=5$~mag with the $\rv=4.0$ extinction law.

Our calculated spectrum for Elias~16 provides a reasonable match to the observed SED over the entire 1--30\mic\ spectral range, allowing for the presence of 8--12\mic\ silicate absorption (Bowey \etal\ 1998) not accounted for in the model. In contrast, the observed SED for IRAS~04262+2735 diverges systematically from the model for wavelengths $\lambda >5$\mic, indicating the presence of strong infrared excess relative to the expected flux from the reddened photosphere in the Rayleigh-Jeans limit. The spectral form of the excess is broadly consistent with emission from dust at temperature $T_{\rm dust}\approx 350$~K, as illustrated in \fig 6. The most probable explanation is that IRAS~04262+2735 is, indeed, a previously unrecognized YSO with a warm circumstellar envelope.

\section{Conclusions}\label{conc}
The main product of this work is the catalog presented in Table~1, which includes 248~confirmed or probable background field stars (and one probable new YSO). This catalog should prove to be a valuable resource for future observing programs. For example, virtually all the included sources are expected to show 3\mic\ \water-ice absorption, based on the previously-known $\av>3$ threshold in the correlation of ice optical depth with extinction (Whittet \etal\ 1988, 2001). The distribution of ice may thus be mapped in more detail than was possible in the previous study of Murakawa \etal\ (2000). The 3\mic\ feature also has potential for mapping magnetic fields within the clouds, by virtue of excess polarization observed at this wavelength when ice-mantled grains are aligned (Hough \etal\ 1988). This method is potentially more reliable than studies utilizing continuum polarization, as the ice feature is an unambiguous tracer of dense material in the line of sight (Whittet \etal\ 2008).

\acknowledgments
This research has made use of the NASA/IPAC Infrared Science Archive, which is operated by the Jet Propulsion Laboratory, California Institute of Technology, under contract with the National Aeronautics and Space Administration (NASA). The Two Micron All Sky Survey is a joint project of the University of Massachusetts and the Infrared Processing and Analysis Center, funded by NASA and the National Science Foundation. The Taurus IRAC Point Source Archive utilizes IRAC data processed from the Taurus Spitzer Legacy Survey (PID~3584, PI~Deborah Padgett). We are grateful to Barbara Whitney and Marilyn Meade for making the unpublished IRAC data available to us. Extensive use was also made of the SIMBAD database, operated at CDS, Strasbourg, France. Financial support for this research was provided by NASA (grant NAG5-12750 and JPL/Caltech Support Agreement no.\ 1264149).

\clearpage
\begin{deluxetable}{llcccccccclr}
\tabletypesize{\scriptsize}
\tablecaption{Catalog of reddened field stars and associated data$^a$\label{tbl-1}} 
\tablewidth{0pt} 
\tablehead{ 
\colhead{2MASS} & \colhead{Other name$^b$} & \colhead{Sp.} & \colhead{$J$--$H$} & \colhead{$H$--$K_{\rm s}$} &
\colhead{$K_{\rm s}$} & \colhead{$[3.6]$} & \colhead{$[4.5]$} & \colhead{$[5.8]$} & \colhead{$[8.0]$} & \colhead{$[12]$} & \colhead{$A_V$}}
\startdata 
J04082673+2803429 & & & 1.17 & 0.50 & 9.60 &&&&&& 6.4\\ 
J04090144+2453214 & & & 1.27 & 0.50 & 5.73 &&&&&& 4.4\\ 
J04092063+2816031 & & & 1.27 & 0.43 & 7.45 &&&&&& 4.1\\
J04104300+2820340 & & & 1.33 & 0.47 & 9.84 &&&&&& 4.6\\
J04110488+2443185 & & & 1.15 & 0.44 & 8.12 &&&&&& 3.5\\
J04112677+2831093 & & & 1.07 & 0.42 & 9.00 &&&&&& 5.5\\
J04112801+2830271 & & & 1.17 & 0.45 & 9.94 &&&&&& 3.7\\
J04113168+2829562 & JH 126 & & 0.90 & 0.40 & 8.74 &&&&&& 4.5\\
J04114185+2841282 & IRAS 04085+2833 & & 1.11 & 0.47 & 5.09 &&&&& 4.60 &5.9\\
J04114938+2815568 & & & 1.16 & 0.45 & 9.24 &&&&&& 3.6\\
J04122296+2949441 & & & 1.09 & 0.43 & 9.99 &&&&&& 5.6\\
J04123318+2945152 & & & 1.32 & 0.53 & 9.25 &&&&&& 4.8\\
J04123940+2816419 & & & 1.64 & 0.62 & 8.85 &&&&&& 7.0\\
J04130664+2235365 & & & 1.03 & 0.40 & 7.53 &&&&&& 5.2\\
J04132688+2804584 & & & 1.11 & 0.52 & 9.54 &&&&&& 6.3\\ 
J04132993+2943325 & & & 1.04 & 0.41 & 9.20 &&&&&& 5.3\\ 
J04133166+2806130 & & & 2.25 & 1.11 & 9.19 &&&&&& 15.2\\ 
J04134374+2821547 & & ~K3\,III$^{\rm d}$ & 1.74 & 0.73 & 7.65 &&&&&& 8.4\\ 
J04134868+2823436 & & & 1.28 & 0.51 & 6.66 &&&&&& 4.5\\ 
J04135352+2813056 & & & 3.90 & 2.06 & 9.92 &&&&&& 28.7\\ 
J04135912+2822182 & & & 1.32 & 0.50 & 9.60 &&&&&& 4.7\\ 
J04140174+2808577 & & & 2.98 & 1.53 & 9.84 &&&&&& 21.2\\ 
J04143434+2802406 & & & 1.29 & 0.59 & 8.97 &&&&&& 7.5\\ 
J04144359+2817086 & & & 2.84 & 1.36 & 8.03 &&&&& 6.83$^{\rm e}$ & 17.0\\ 
%J04145234+2805598 & & & 1.32 & 0.50 & 7.71 &&&&&& 4.7\\ % X-ray YSO
J04152407+2807074 & & & 0.88 & 0.44 & 9.18 &&&&&& 6.3\\ 
J04152859+2451554 & IRAS 04124+2444 & & 1.03 & 0.43 & 4.74 &&&&& 4.49 & 5.3\\ 
J04152950+2818313 & & & 0.95 & 0.41 & 9.70 &&&&&& 4.8\\ 
J04154949+2851175 & & & 1.39 & 0.60 & 9.82 &&&&&& 5.5\\ 
J04162595+2808443 & & & 1.31 & 0.51 & 8.66 &&&&&& 4.6\\ 
J04163376+2854051 & & & 0.91 & 0.41 & 8.80 &&&&&& 4.6\\ 
J04163442+2802386 & JH 167 & & 1.08 & 0.42 & 7.20 &&&&&& 3.0\\ 
J04163846+2853573 & & & 1.04 & 0.40 & 9.50 &&&&&& 2.7\\ 
J04170129+2839143 & & & 1.96 & 0.97 & 9.65 &&&&&& 12.9\\ 
J04170178+2821593 & & & 1.79 & 0.76 & 8.53 & 8.07 & 8.13 & 7.93 & 7.95 && 8.5\\ 
J04173477+2757338 & & & 1.37 & 0.63 & 9.78 & 9.55 & 9.40 & 9.31 & 9.29 && 8.2\\ 
J04173746+2811230 & & & 1.82 & 0.76 & 9.35 & 8.90 & 8.85 & 8.73 & 8.74 && 8.6\\ 
J04174322+2747396 & & & 1.14 & 0.48 & 9.78 & 9.49 & 9.49 & 9.40 & 9.39 && 6.2\\ 
J04180306+2840528 & & & 0.79 & 0.46 & 9.87 &      &      &      &      && 5.9\\ 
J04181078+2519574 & & & 1.16 & 0.56 & 9.03 &      &      &      &      && 6.6\\ 
%J04182909+2826191 & & & 3.27 & 1.70 & 9.94 & 8.87 & 8.53 & 8.36 & 8.35 && 23.6\\  % Luhman YSO
J04183702+2434105 & & & 1.09 & 0.42 & 6.67 &      &      &      &      && 3.0\\ 
J04184535+2826400 & V410 Anon 9 & A2 & 1.49 & 0.81 & 7.91 & 7.33 & 7.26 & 7.14 & 7.14 && 11.8\\ 
J04184767+2834011 & & & 1.19 & 0.47 & 9.65 &      &      &      &      && 3.9\\ 
J04192736+2813012 & & & 1.55 & 0.58 & 8.16 & 7.83 & 7.96 & 7.72 & 7.71 && 6.3\\ 
J04195830+2812139 & & & 1.32 & 0.46 & 7.50 & 7.15 & 7.23 & 7.09 & 7.10 && 4.5\\ 
J04203249+2721322 & & & 1.03 & 0.41 & 8.49 & 8.24 & 8.40 & 8.23 & 8.21 && 5.3\\ 
J04203895+2706404 & & & 1.38 & 0.60 & 8.96 &      &      &      &      && 5.5\\ 
J04204138+2705474 & & & 1.28 & 0.55 & 7.68 &      &      &      &      && 7.3\\ 
J04205771+2516184 & & & 1.27 & 0.49 & 8.63 &      &      &      &      && 4.3\\ 
J04205996+3022179 & & & 0.96 & 0.40 & 6.22 &      &      &      &      && 4.8\\ 
J04211454+2703302 & & & 1.63 & 0.72 & 9.21 &      &      &      &      && 7.4\\ 
J04213256+2657270 & & & 1.69 & 0.78 & 9.24 &      &      &      &      && 8.1\\ 
J04214879+2832553 & & & 1.25 & 0.50 & 8.21 & 7.83 & 7.89 & 7.79 & 7.76 && 4.3\\ 
%J04222404+2646258 & GJL 0419.3+2639 & & 0.90 & 0.42 & 9.77 &      &      &      &      && 5.5\\ % X-ray YSO?
J04223539+2504186 & & & 1.16 & 0.43 & 9.43 &      &      &      &      && 3.5\\ 
J04231777+2806260 & & & 1.08 & 0.46 & 6.52 & 6.30 & 6.43 & 6.26 & 6.20 && 5.8\\ 
J04231783+2508529 & & & 0.95 & 0.46 & 9.63 &      &      &      &      && 6.7\\ 
J04232455+2500084 & Elias 3 & K2\,III & 1.74 & 0.78 & 5.81 &      &      &      &      && 9.2\\ 
J04234626+2642457 & GJL 0420.7+2636 & ~K5\,III$^{\rm d}$ & 1.86 & 0.80 & 9.31 & 8.83 & 8.73 & 8.61 & 8.56 && 8.8\\ 
J04251866+2555359 & & & 1.04 & 0.44 & 6.82 & 6.52 & 6.55 & 6.31 & 6.19 && 5.4\\ 
J04253953+2544592 & IRAS 04225+2538 & & 1.04 & 0.41 & 5.21 & ---  & ---  & 4.99 & 4.94 & 5.06 & 5.3\\ 
J04254287+2635535 & & & 1.08 & 0.41 & 9.79 & 9.52 & 9.52 & 9.37 & 9.43 && 2.9\\ 
J04255543+2709122 & & & 1.12 & 0.46 & 8.01 & 7.74 & 7.73 & 7.62 & 7.65 && 5.9\\ 
J04261745+2436588 & & & 1.65 & 0.68 & 9.27 & 8.92 & 8.91 & 8.73 & 8.77 && 7.3\\ 
J04262182+2652111 & & & 1.25 & 0.41 & 8.46 & 8.19 & 8.28 & 8.17 & 8.13 && 3.8\\ 
J04263071+2436372 & & & 2.85 & 1.38 & 7.62 & 6.79 & 6.74 & 6.48 & 6.39 & 6.41$^{\rm e}$& 17.8\\ 
J04263650+2439469 & & & 2.27 & 1.11 & 9.59 &      &      &      &      && 15.3\\ 
J04274283+2622567 & & & 1.44 & 0.58 & 7.36 & 6.98 & 7.14 & 6.95 & 6.94 && 5.7\\ 
J04274738+2442267 & & & 1.16 & 0.45 & 9.98 & 9.67 & 9.61 & 9.53 & 9.54 && 3.6\\ 
J04274755+2624181 & & & 1.21 & 0.41 & 8.32 & 8.01 & 8.06 & 7.96 & 7.92 && 3.6\\ 
J04275374+2617222 & & & 1.13 & 0.43 & 9.62 & 9.32 & 9.29 & 9.19 & 9.19 && 3.3\\ 
J04275776+2440599 & & & 1.63 & 0.68 & 8.76 & 8.37 & 8.33 & 8.13 & 8.12 && 7.2\\ 
J04280463+2438411 & & & 1.34 & 0.56 & 9.11 & 8.82 & 8.77 & 8.67 & 8.61 && 5.1\\ 
J04280994+2432059 & & & 1.13 & 0.42 & 8.84 & 8.55 & 8.55 & 8.42 & 8.47 && 3.3\\ 
J04281255+2440431 & & & 1.25 & 0.55 & 9.01 & 8.79 & 8.71 & 8.62 & 8.59 && 7.1\\ 
J04282062+2653387 & & & 1.33 & 0.51 & 7.75 & 7.35 & 7.44 & 7.30 & 7.30 && 4.8\\ 
J04285007+2438275 & & & 1.30 & 0.52 & 9.12 & 8.80 & 8.75 & 8.64 & 8.65 && 4.7\\ 
J04285165+2433477 & & & 1.23 & 0.52 & 8.76 & 8.40 & 8.35 & 8.34 & 8.29 && 6.8\\ 
J04290729+2659136 & & & 1.27 & 0.48 & 9.18 & 8.85 & 8.86 & 8.76 & ---  && 4.3\\ 
J04291286+2442532 & & & 1.54 & 0.65 & 9.41 &      &      &      &      && 6.6\\ 
J04292083+2742074 & IRAS 04262+2735 & YSO?$^{\rm \,f}$ & 1.09 & 0.45 & 7.02 & 6.55 & 6.50 & 6.15 & 5.01 & 3.90 & 5.0\\
J04293024+2723521 & & & 1.29 & 0.46 & 7.32 & 6.92 & 7.22 & 7.01 & 7.01 && 4.3\\ 
J04293024+2658276 & & & 2.26 & 1.02 & 8.13 & 7.45 & 7.57 & 7.27 & 7.27 & 6.96$^{\rm e}$& 11.5\\ 
J04294376+2701532 & & & 1.37 & 0.51 & 8.74 & 8.40 & 8.43 & 8.30 & 8.32 && 5.0\\ 
J04294411+2438446 & & & 1.21 & 0.41 & 8.36 & 8.18 & 8.27 & 8.10 & 8.09 && 3.6\\ 
J04294651+2431493 & & & 1.81 & 0.77 & 9.55 & 9.12 & 8.92 & 8.80 & 8.81 && 8.6\\ 
J04295364+2332364 & & & 1.27 & 0.50 & 6.26 & ---  & 6.11 & 5.93 & 5.89 && 4.4\\ 
J04295531+2258579 & IRAS 04269+2252 & & 1.14 & 0.53 & 4.72 & ---  & ---  & 4.35 & 4.18 & 3.49 & 6.5\\ 
J04301480+2717460 & & & 1.15 & 0.41 & 8.59 & 8.37 & 8.43 & 8.33 & 8.32 && 3.3\\ 
J04302414+2819165 & & & 1.20 & 0.44 & 7.14 & 6.80 & 6.85 & 6.65 & 6.54 && 3.7\\ 
J04303410+2711046 & & & 1.25 & 0.44 & 8.21 & 7.89 & 8.01 & 7.87 & 7.82 && 4.0\\ 
J04303867+2255520 & & & 1.55 & 0.77 & 7.74 & 7.31 & 7.27 & 7.14 & 7.12 && 9.8\\ 
J04304246+2258248 & & & 1.21 & 0.53 & 8.14 & 7.76 & 7.83 & 7.70 & 7.67 && 6.8\\ 
J04304284+2743299 & & & 1.05 & 0.52 & 7.67 & 6.92 & 6.97 & 6.71 & 6.62 && 7.6\\ 
J04305298+2409541 & & & 1.23 & 0.48 & 9.83 & 9.52 & 9.46 & 9.41 & 9.39 && 4.1\\ 
J04305335+2709552 & & & 1.08 & 0.41 & 9.57 & 9.27 & 9.29 & 9.22 & 9.21 && 3.0\\ 
J04305639+2409078 & & & 1.21 & 0.51 & 8.40 & 8.12 & 8.14 & 8.02 & 8.00 && 6.7\\ 
J04312113+2658422 & IRAS 04282+2652 & & 1.23 & 0.63 & 4.91 & ---  & --- & 4.36 & 4.19 & 3.86 & 7.4\\ 
J04312636+2707204 & JH 57 &  & 0.92 & 0.45 & 8.23 & 8.05 & 7.93 & 7.84 & 7.85 && 6.5\\ 
J04313158+2439424 & & & 1.09 & 0.44 & 6.41 & 6.18 & 6.32 & 6.18 & 6.14 && 5.7\\ 
J04314179+2440149 & & & 1.33 & 0.48 & 8.93 & 8.65 & 8.78 & 8.58 & 8.57 && 4.7\\ 
J04314202+2704049 & & & 1.16 & 0.47 & 9.43 & 9.16 & 9.15 & 9.06 & 9.00 && 6.2\\ 
J04314783+2415023 & & & 1.53 & 0.59 & 9.62 & 9.20 & 9.28 & 9.09 & 9.10 && 6.2\\ 
J04314862+2706170 & & & 1.04 & 0.48 & 9.26 & 8.95 & 8.92 & 8.82 & 8.86 && 5.6\\ 
J04315067+2659412 & JH 64 & & 1.00 & 0.41 & 8.89 & 8.62 & 8.76 & 8.72 & 8.64 && 5.0\\ 
J04315117+2443449 & & & 1.32 & 0.45 & 7.99 & 7.68 & 7.80 & 7.67 & 7.62 && 4.4\\ 
J04315860+2412552 & & & 1.15 & 0.44 & 9.12 & 8.86 & 8.91 & 8.77 & 8.77 && 3.5\\ 
J04320127+2813347 & & & 1.15 & 0.40 & 6.89 & 6.63 & 6.82 & 6.64 & 6.63 && 3.3\\ 
J04320792+2432106 & & & 1.41 & 0.54 & 7.97 & 7.59 & 7.63 & 7.50 & 7.47 && 5.4\\ 
J04320816+2405482 & & & 1.18 & 0.43 & 9.79 & 9.49 & 9.54 & 9.43 & 9.40 && 3.6\\ 
J04321130+2613237 & & & 1.09 & 0.41 & 9.23 & 8.94 & 9.01 & 8.89 & 8.83 && 3.0\\ 
J04321153+2433380 & Elias 9 & M4\,III & 1.53 & 0.58 & 5.48 & ---  & ---  & 5.06 & 5.01 & 5.22$^{\rm e}$ & 4.9\\ 
J04321327+2429107 & & & 3.46 & 1.78 & 8.12 & 7.06 & 6.77 & 6.54 & 6.52 & 6.60$^{\rm e}$ & 22.5\\ 
J04321380+2630461 & & & 1.30 & 0.45 & 9.26 & 8.94 & 9.00 & 8.90 & 8.84 && 4.3\\ 
J04322815+2711228 & & & 1.44 & 0.60 & 7.16 & 6.70 & 6.77 & 6.53 & 6.40 && 5.8\\ 
J04323594+2253112 & & & 1.31 & 0.54 & 7.05 &      &      &      &      && 4.8\\ 
J04323842+2648098 & & & 1.21 & 0.42 & 8.88 & 8.62 & 8.67 & 8.60 & 8.57 && 3.7\\ 
J04323892+2358251 & Elias 10 & M8\,III & 1.17 & 0.54 & 5.81 & ---  & ---  & 4.83 & 4.65 & 4.49 & 2.2\\ 
J04323908+2700083 & IRAS 04295+2653 & & 1.03 & 0.41 & 4.89 & ---  & ---  & 4.69 & 4.64 & 4.74 & 5.2\\
J04324166+2419038 & & & 1.19 & 0.59 & 8.88 & 8.58 & 8.36 & 8.35 & 8.35 && 7.0\\ 
J04325815+2525324 & & & 1.11 & 0.43 & 8.00 & 7.81 & 7.95 & 7.78 & 7.73 && 3.2\\ 
J04330802+2556436 & & & 1.50 & 0.65 & 8.65 & 8.19 & 8.19 & 8.05 & 8.04 && 6.4\\ 
J04330971+2656220 & & ~K8\,III$^{\rm d}$ & 1.41 & 0.54 & 6.85 & 6.49 & 6.61 & 6.45 & 6.41 && 5.0\\ 
J04332164+2239504 & JH 114 & & 1.10 & 0.42 & 7.19 &      &      &      &      && 3.1\\ 
J04332594+2615334 & Elias 13 & K2\,III & 2.03 & 0.97 & 5.56 & ---  & ---  & 4.78 & 4.76 & 4.90 & 11.7\\ 
%J04332621+2245293 & & & 1.30 & 0.58 & 9.92 &      &      &      &      && 7.5\\ % X-ray YSO?
J04332662+2410446 & & & 1.17 & 0.41 & 8.59 & 8.35 & 8.41 & 8.29 & 8.25 && 3.4\\ 
J04333234+2425120 & & & 1.22 & 0.45 & 8.13 & 7.72 & 7.82 & 7.75 & 7.70 && 3.9\\ 
J04333734+2247505 & & & 1.35 & 0.50 & 8.12 &      &      &      &      && 4.8\\ 
J04333776+2612591 & & & 1.06 & 0.57 & 9.33 & 8.95 & 8.83 & 8.76 & 8.78 && 7.9\\ 
J04334115+2408100 & & & 1.20 & 0.48 & 9.94 & 9.64 & 9.59 & 9.53 & 9.49 && 4.0\\ 
J04334465+2615005 & & & 1.25 & 0.64 & 9.74 &      &      &      &      && 9.3\\ 
J04335006+2408216 & & & 1.11 & 0.46 & 9.21 & 8.91 & 8.95 & 8.82 & 8.82 && 5.9\\ 
J04335113+2615115 & & & 1.17 & 0.48 & 8.43 & 8.05 & 8.03 & 7.99 & 7.94 && 3.8\\ 
J04335142+2718339 & & & 1.18 & 0.44 & 8.63 & 8.28 & 8.51 & 8.36 & 8.31 && 3.6\\ 
J04340563+2734335 & & & 1.24 & 0.49 & 6.09 & ---  & 6.00 & 5.80 & 5.78 && 4.2\\ 
J04341640+2651195 & & & 1.34 & 0.51 & 9.21 & 8.89 & 8.95 & 8.85 & 8.85 && 4.8\\ 
J04342761+2706433 & & & 1.32 & 0.49 & 9.86 & 9.50 & 9.54 & 9.43 & 9.40 && 4.6\\ 
J04343077+2541497 & & & 1.26 & 0.43 & 7.75 & 7.45 & 7.60 & 7.45 & 7.44 && 4.0\\ 
J04343120+2653363 & & & 0.73 & 0.42 & 9.79 & 9.49 & 9.46 & 9.41 & 9.44 && 5.4\\ 
J04343230+2250218 & & & 1.11 & 0.42 & 9.05 &      &      &      &      && 3.1\\ 
J04343549+2644062 & & & 1.30 & 0.52 & 6.89 & 6.49 & 6.58 & 6.43 & 6.20 && 4.7\\ 
J04343848+2242133 & & & 1.05 & 0.44 & 6.89 &      &      &      &      && 5.5\\ 
J04344389+2655155 & & & 1.25 & 0.49 & 9.55 & 9.21 & 9.20 & 9.08 & 9.07 && 4.2\\ 
J04344754+2657083 & & & 1.42 & 0.49 & 9.03 & 8.73 & 8.75 & 8.63 & 8.61 && 5.1\\ 
J04351798+2402459 & & & 1.15 & 0.53 & 9.15 & 8.84 & 8.85 & 8.75 & 8.73 && 6.5\\ 
J04352020+2232146 & & & 0.96 & 0.51 & 9.73 &      &      &      &      && 7.1\\ 
J04352271+2647297 & & & 1.21 & 0.43 & 9.54 & 9.21 & 9.27 & 9.16 & 9.15 && 3.7\\ 
J04352370+2404502 & & & 2.40 & 1.25 & 8.51 & 7.75 & 7.65 & 7.43 & 7.42 && 16.7\\ 
J04352820+2250257 & & & 1.09 & 0.41 & 8.13 &      &      &      &      && 3.0\\ 
J04353751+2405332 & & & 2.00 & 0.95 & 9.37 & 8.81 & 8.78 & 8.58 & 8.57 && 10.6\\ 
J04353776+2403348 & & & 1.61 & 0.67 & 9.29 & 8.93 & 8.91 & 8.77 & 8.78 && 7.1\\ 
J04354848+2251341 & & & 1.29 & 0.46 & 9.43 &      &      &      &      && 4.3\\ 
%J04355209+2255039 & & & 1.08 & 0.42 & 9.81 &      &      &      &      && 3.0\\ % X-ray YSO?
J04355684+2254360 & & & 1.09 & 0.52 & 9.53 &      &      &      &      && 6.1\\ 
J04361252+2413387 & & & 1.13 & 0.42 & 9.79 & 9.56 & 9.64 & 9.53 & 9.52 && 3.3\\ 
J04363003+2318383 & & & 1.17 & 0.48 & 6.79 & 6.48 & 6.72 & 6.52 & 6.49 && 6.3\\ 
J04363513+2526425 & & & 1.08 & 0.43 & 7.12 & 6.83 & 6.92 & 6.80 & 6.78 && 5.6\\ 
J04364224+2555117 & IRAS 04336+2549 & & 1.17 & 0.40 & 5.26 & ---  & ---  & 4.74 & 4.65 & 4.72 & 3.4\\ 
J04365561+2336594 & & & 1.23 & 0.51 & 6.80 & 6.20 & 6.36 & 6.33 & 5.98 && 4.3\\ 
J04365910+2619418 & & & 1.48 & 0.52 & 6.76 & 6.38 & 6.52 & 6.36 & 6.34 && 5.6\\ 
J04371368+2422208 & & & 1.11 & 0.42 & 6.86 & 6.50 & 6.73 & 6.58 & 6.57 && 3.2\\ 
J04371558+2616155 & & & 1.29 & 0.45 & 8.11 & 7.75 & 7.84 & 7.72 & 7.69 && 4.3\\ 
J04371580+2629295 & & & 1.89 & 0.88 & 9.94 & 9.52 & 9.34 & 9.18 & 9.15 && 9.6\\ 
J04371735+2255406 & & & 1.02 & 0.41 & 8.74 &      &      &      &      && 5.2\\ 
J04372471+2627285 & & & 1.88 & 0.79 & 7.62 & 7.10 & 7.20 & 6.99 & 6.98 && 9.1\\ 
J04372821+2610289 & TNS 2; Kim 6 & ~M0\,III$^{\rm c}$ & 1.63 & 0.67 & 6.70 & 6.29 & 6.40 & 6.19 & 6.17 & 6.04$^{\rm e}$ & 6.6\\ 
J04372946+2609509 & & & 1.08 & 0.43 & 9.91 & 9.62 & 9.58 & 9.51 & 9.50 && 5.6\\ 
J04373457+2625518 & & & 1.15 & 0.41 & 8.37 & 8.14 & 8.24 & 8.06 & 8.01 && 3.3\\ 
J04375136+2623585 & & & 1.09 & 0.41 & 7.13 & 6.88 & 6.96 & 6.82 & 6.81 && 3.0\\ 
J04375850+2602497 & & & 1.32 & 0.49 & 9.65 & 9.44 & 9.35 & 9.28 & 9.27 && 4.6\\ 
J04375986+2528090 & IRAS 04349+2522 & ~M3\,III$^{\rm c}$ & 1.21 & 0.50 & 5.20 & ---  & ---  & 4.90 & 4.89 & 4.78 & 2.9\\ 
J04383741+2547138 & & & 1.10 & 0.42 & 9.66 & 9.36 & 9.41 & 9.31 & 9.29 && 3.1\\ 
J04383928+2551062 & Kim 23 & ~K1\,III$^{\rm c}$ & 1.35 & 0.57 & 9.13 & 8.71 & 8.73 & 8.65 & 8.65 && 6.4\\ 
J04383931+2608508 & & & 1.30 & 0.55 & 9.96 & 9.64 & 9.67 & 9.55 & 9.50 && 4.8\\ 
J04383974+2619310 & & & 1.17 & 0.41 & 9.57 & 9.27 & 9.34 & 9.24 & 9.20 && 3.4\\ 
J04384018+2639382 & & & 1.55 & 0.66 & 7.30 & 6.88 & 6.92 & 6.73 & 6.72 && 6.7\\ 
J04384470+2518004 & & & 1.27 & 0.43 & 8.12 & 7.87 & 7.94 & 7.77 & 7.74 && 4.1\\ 
J04385148+2534172 & TNS 10; Kim 30 & ~M0\,III$^{\rm c}$ & 1.21 & 0.45 & 4.87 & ---  & ---  & 4.59 & 4.59 & 4.58 & 3.3\\
J04385153+2559441 & Kim 29 & ~K3\,III$^{\rm c}$ & 1.29 & 0.50 & 7.38 & 7.04 & 7.10 & 6.98 & 6.98 && 5.0\\ 
J04385827+2631084 & Elias 14 & M5\,III & 1.72 & 0.80 & 6.89 & 6.35 & 6.37 & 6.15 & 5.98 && 6.7\\ 
J04390593+2550079 & Kim 32 & ~K0\,III$^{\rm c}$ & 1.73 & 0.75 & 8.17 & 7.68 & 7.64 & 7.48 & 7.49 && 8.1\\ 
J04390696+2627199 & JH 214 & & 0.95 & 0.48 & 9.09 & 8.73 & 8.64 & 8.60 & 8.65 && 9.6\\ 
J04390745+2553544 & & & 3.10 & 1.50 & 9.97 & 9.00 & 8.85 & 8.61 & 8.58 && 19.1\\ 
J04390885+2614103 & & & 1.38 & 0.59 & 9.84 & 9.50 & 9.46 & 9.39 & 9.33 && 5.5\\ 
J04391820+2544326 & & & 1.58 & 0.63 & 9.53 & 9.06 & 9.11 & 8.97 & 8.98 && 6.7\\ 
J04392692+2552592 & Elias 15 & M2\,III & 2.75 & 1.28 & 6.91 & 6.08 & 6.10 & 5.81 & 5.79 & 5.93$^{\rm e}$ & 15.3\\ 
J04392986+2618317 & & & 1.13 & 0.49 & 9.97 & 9.64 & 9.61 & 9.55 & 9.50 && 6.2\\ 
J04393558+2628396 & & & 1.01 & 0.44 & 9.82 & 9.52 & 9.45 & 9.38 & 9.43 && 5.3\\ 
J04393779+2613567 & & & 1.49 & 0.64 & 9.68 & 9.28 & 9.27 & 8.99 & 9.10 && 6.3\\ 
J04393890+2611250 & Elias 16 & K1\,III & 3.64 & 1.81 & 5.16 & ---  & ---  & 3.56 & 3.54 & 3.72 & 24.1\\ 
J04394383+2535450 & & & 1.70 & 0.66 & 9.98 & 9.59 & 9.52 & 9.39 & 9.38 && 7.5\\ 
J04394395+2516013 & Elias 17 & M5\,III & 1.44 & 0.62 & 6.56 & 6.18 & 6.40 & 6.14 & 6.15 && 4.3\\ 
J04400300+2528176 & & & 1.50 & 0.64 & 9.42 & 8.98 & 9.00 & 8.85 & 8.83 && 6.4\\ 
J04401230+2613203 & JH 216; Kim 45 & ~G8\,III$^{\rm c}$ & 1.07 & 0.43 & 8.30 & 7.98 & 8.01 & 7.89 & 7.92 && 4.7\\ 
J04401383+2559163 & Kim 46 & ~K5\,III$^{\rm c}$ & 1.79 & 0.71 & 7.44 & 7.01 & 7.03 & 6.84 & 6.85 && 8.0\\ 
J04403830+2554274 & GKH 7  & & 1.05 & 0.42 & 9.06 &      &      &      &      && 5.3\\ 
J04404541+2531566 & Kim 49 & ~G4\,III$^{\rm c}$ & 1.94 & 0.84 & 9.33 &      &      &      &      && 11.6\\ 
J04405597+2531312 & Kim 52 & ~K2\,III$^{\rm c}$ & 1.61 & 0.70 & 7.22 &      &      &      &      && 8.1\\ 
J04405690+2601043 & & & 1.76 & 0.72 & 8.76 & 8.34 & 8.43 & 8.17 & 8.14 && 8.1\\ 
J04405745+2554134 & TNS 8 & K5\,III & 3.37 & 1.71 & 7.46 & 6.38 & 6.17 & 5.94 & 5.87 && 21.5\\ 
J04405845+2612306 & & & 1.20 & 0.48 & 9.80 & 9.45 & 9.50 & 9.44 & 9.37 && 3.9\\ 
J04410274+2539470 & TNS 12 & & 1.33 & 0.58 & 5.20 & ---  & ---  & 4.83 & 4.80 & 4.95 & 6.9\\ 
J04410544+2541078 & & & 1.12 & 0.42 & 9.66 & 9.33 & 9.37 & 9.27 & 9.25 && 3.2\\ 
J04411845+2401576 & & & 1.12 & 0.43 & 6.58 & 6.28 & 6.28 & 6.11 & 5.94 && 3.2\\ 
J04412285+2556199 & & & 1.03 & 0.48 & 9.58 & 9.28 & 9.16 & 9.11 & 9.08 && 5.6\\ 
J04413015+2527019 & Kim 59 & ~K2\,III$^{\rm c}$ & 2.01 & 0.89 & 7.63 & 6.99 & 7.02 & 6.85 & 6.83 && 11.2\\ 
J04413108+2519222 & & & 1.85 & 0.79 & 9.50 & 8.97 & 8.97 & 8.77 & 8.73 && 8.9\\ 
J04413486+2500544 & & & 1.54 & 0.65 & 9.24 & 8.79 & 8.74 & 8.60 & 8.57 && 6.6\\ 
J04413630+2502062 & & & 1.29 & 0.51 & 8.94 & 8.55 & 8.57 & 8.48 & 8.45 && 4.6\\ 
J04413973+2505583 & & & 1.25 & 0.41 & 8.36 & 8.06 & 8.19 & 8.02 & 7.98 && 3.8\\ 
J04415119+2602090 & & & 1.10 & 0.40 & 9.43 & 9.20 & 9.31 & 9.20 & 9.16 && 3.0\\ 
J04415321+2449158 & & & 1.16 & 0.44 & 7.82 & 7.58 & 7.67 & 7.47 & 7.41 && 3.5\\ 
J04415533+2515092 & & & 1.04 & 0.41 & 9.67 & 9.43 & 9.35 & 9.30 & 9.25 && 5.3\\ 
J04415802+2600435 & & & 1.17 & 0.42 & 7.56 & 7.26 & 7.45 & 7.26 & 7.28 && 3.5\\ 
J04420337+2540258 & JH 218 & & 0.92 & 0.42 & 8.90 & 8.59 & 8.56 & 8.52 & 8.54 && 4.7\\ 
J04420618+2543246 & & & 1.10 & 0.41 & 9.79 & 9.57 & 9.54 & 9.44 & 9.43 && 3.1\\ 
J04421818+2536207 & & & 1.43 & 0.60 & 9.83 & 9.42 & 9.42 & 9.30 & 9.25 && 5.8\\ 
J04421841+2516532 & Kim 65 & ~K1\,III$^{\rm c}$ & 1.06 & 0.44 & 8.68 & 8.38 & 8.39 & 8.29 & 8.25 && 4.3\\ 
J04422934+2517425 & GKH 21 & & 1.25 & 0.52 & 8.86 & 8.54 & 8.60 & 8.46 & 8.46 && 4.4\\ 
J04423570+2527152 & IRAS 04395+2521 & ~K5\,III$^{\rm c}$ & 1.27 & 0.52 & 5.50 & --- & --- & 5.10 & 5.13 & 5.12 & 4.4\\% Kim 69
J04424107+2505064 & & & 1.37 & 0.52 & 9.61 & 9.22 & 9.34 & 9.23 & 9.21 && 5.0\\ 
J04425915+2507206 & & & 1.10 & 0.42 & 9.11 & 8.80 & 8.79 & 8.72 & 8.68 && 3.1\\ 
J04430048+2459230 & IRAS 04399+2453 & & 1.12 & 0.41 & 5.48 & ---  & --- & 5.18 & 5.19 & 5.15 & 3.2\\ 
J04430889+2504490 & & & 1.13 & 0.43 & 9.62 & 9.31 & 9.42 & 9.28 & 9.27 && 3.3\\ 
J04431258+2521027 & GKH 25 & & 1.22 & 0.43 & 9.00 & 8.76 & 8.76 & 8.64 & 8.65 && 3.8\\ 
J04431538+2444558 & & & 1.27 & 0.54 & 6.80 & 6.49 & 6.65 & 6.46 & 6.39 && 4.6\\ 
J04434337+2523257 & GKH 27; Kim 84 & & 1.11 & 0.41 & 7.33 & 7.04 & 7.15 & 7.02 & 7.02 && 3.1\\ 
J04434870+2457306 & & & 1.15 & 0.46 & 6.14 & ---  & 6.06 & 5.90 & 5.84 && 3.6\\ 
J04441137+2459355 & & & 1.22 & 0.43 & 8.43 & 8.13 & 8.31 & 8.14 & 8.10 && 3.8\\ 
J04441794+2524512 & Elias 19 & ~M4\,III~ & 1.18 & 0.46 & 6.05 & ---  & 5.94 & 5.71 & 5.65 && 2.5\\ 
J04443504+2501082 & IRAS 04415+2455 & & 1.28 & 0.52 & 5.97 & ---  & 5.66 & 5.42 & 5.29 & 5.25 & 4.6\\ 
J04445592+2509204 & & & 1.15 & 0.40 & 7.92 & 7.73 & 7.79 & 7.66 & 7.64 && 3.3\\ 
J04450980+2503186 & & & 1.31 & 0.45 & 7.76 & 7.47 & 7.59 & 7.42 & 7.43 && 4.3\\ 
J04451569+2449525 & & & 1.22 & 0.41 & 6.58 & 6.31 & 6.42 & 6.29 & 6.26 && 3.7\\ 
J04451757+2446027 & & & 1.16 & 0.42 & 6.01 & ---  & 5.92 & 5.76 & 5.70 && 3.4\\ 
J04452612+2501590 & & & 1.13 & 0.44 & 9.77 & 9.48 & 9.49 & 9.42 & 9.37 && 3.4\\ 
J04453986+2517045 & & & 1.18 & 0.45 & 8.46 & 8.19 & 8.25 & 8.05 & 7.94 && 3.7\\ 
J04461799+2507138 & & & 1.16 & 0.41 & 7.77 & 7.49 & 7.65 & 7.50 & 7.46 && 3.4\\ 
J04463986+2425260 & IRAS 04435+2419 & & 1.13 & 0.40 & 5.87 & ---  & 5.55 & 5.35 & 5.22 & 4.98 & 3.2\\ 
J04471909+2448510 & & & 1.23 & 0.41 & 7.08 & 6.85 & 6.97 & 6.81 & 6.79 && 3.7\\ 
J04472659+2605172 & & & 1.07 & 0.40 & 5.63 & ---  & 5.65 & 5.42 & 5.42 && 2.9\\ 
J04475068+2529299 & & & 1.19 & 0.41 & 9.56 & 9.26 & 9.30 & 9.22 & 9.19 && 3.5\\ 
J04475454+2523191 & & & 1.24 & 0.42 & 9.65 & 9.37 & 9.48 & 9.35 & 9.31 && 3.8\\ 
J04475943+2537551 & & & 1.04 & 0.44 & 9.83 & 9.62 & 9.53 & 9.34 & 9.46 && 5.4\\ 
J04480327+2540079 & & & 1.09 & 0.44 & 9.69 & 9.40 & 9.38 & 9.34 & 9.30 && 5.7\\ 
J04480380+2521488 & GKH 30 & & 1.27 & 0.46 & 7.89 & 7.53 & 7.75 & 7.55 & 7.54 && 4.2\\ 
J04481020+2537537 & & & 1.17 & 0.45 & 8.95 & 8.60 & 8.54 & 8.57 & 8.51 && 3.6\\ 
J04481911+2545075 & GKH 15 & & 1.15 & 0.42 & 7.03 & 6.73 & 6.90 & 6.71 & 6.73 && 3.4\\ 
J04482478+2524559 & & & 1.41 & 0.51 & 7.83 & 7.46 & 7.59 & 7.40 & 7.35 && 5.2\\ 
J04484817+2538123 & Elias 21 & M2\,III & 1.21 & 0.46 & 6.21 & --- & 6.11 & 5.95 & 5.88 && 3.1\\ 
J04494966+2538435 & & & 1.14 & 0.41 & 7.32 & 7.12 & 7.27 & 7.08 & 7.06 && 3.3\\ 
J04512418+2520029 & IRAS 04483+2514 & & 1.12 & 0.50 & 5.53 &&&& & 4.72 & 6.1\\ 
\enddata 
\tablenotetext{a}{Spectral classifications (col.\,3) are from SIMBAD unless otherwise noted. 
Photometry in the $JH\kmag$ passbands (cols.\,4--6) is from the 2MASS catalog; photometry in the
3.6--8.0\mic\ passbands (cols.\,7--10) is from Spitzer IRAC observations; photometry at 12\mic\ 
(col.\,11) is calculated from flux data in the IRAS Point Source Catalog (version~2.0) unless 
otherwise noted; dashes in the photometry columns indicate saturation, empty fields indicate 
no data available. Visual extinction estimates from the current work are listed in the final column.}
\tablenotetext{b}{Key to identifications: Elias -- Elias 1978; GJL -- Goodman et~al. 1992; 
	GKH -- Gomez, Kenyon, \& Hartmann 1994; IRAS~-- Infrared Astronomical Satellite Point Source Catalog; 
	JH -- Jones \& Herbig 1979; Kim -- unpublished catalog cited by Murakawa et~al.\ 2000; TNS -- Tamura et~al.\ 1987; 
	V410 Anon 9 -- Strom \& Strom 1994.}
\tablenotetext{c}{Spectral classification from Murakawa et~al.\ 2000.}
\tablenotetext{d}{Spectral classification from Shenoy 2003.}
\tablenotetext{e}{Photometry simulated from Spitzer IRS 9.9--19.6\mic\ calibrated flux spectra (Whittet \etal\ 2007).}
\tablenotetext{f}{Excess infrared flux at $\lambda > 5$\mic\ suggests this star to be a YSO (see \S5).}
\end{deluxetable}

\clearpage

\begin{deluxetable}{llcc} 
\tabletypesize{\scriptsize} 
\tablecaption{YSOs with high estimated extinction$^a$\label{tbl-2}} 
\tablewidth{0pt} 
\tablehead{ 
\colhead{2MASS} & \colhead{Other name} & \colhead{Sp$^b$} & \colhead{$A_V$}}
\startdata 
J04182239+2824375 & V410 Anon 24 &     & 18.9 \\
J04182909+2826191 & V410 Anon 25 &     & 21.8 \\
J04183444+2830302 & V410 X-ray 2 &     & 19.6 \\
J04184023+2824245 & V410 X-ray 4 &     & 16.4 \\
J04184133+2827250 & LR1          &     & 22.8 \\
J04184505+2820528 & V410 Anon 20 &     & 19.1 \\
J04323205+2257266 & IRAS 04295+2251 &  & 18.5 \\
J04395574+2545020 & Elias 18; IC 2087 IR & B5 & 22.2 \\
J04400800+2605253 & IRAS 04370+2559 &  & 12.5 \\
J04412464+2543530 & ITG 40 &      M3.5 & 21.7 \\
\enddata
\tablenotetext{a}{Based on photometry listed in Luhman \etal\ 2006.}
\tablenotetext{b}{Spectral types are from SIMBAD (Elias~18) and Luhman \etal\ 2006 (ITG~40); intrinsic colors appropriate to an M0 dwarf are assumed in all other cases.}
\end{deluxetable}

\clearpage

\begin{figure}
\centering
\includegraphics[width=14cm, angle=0]{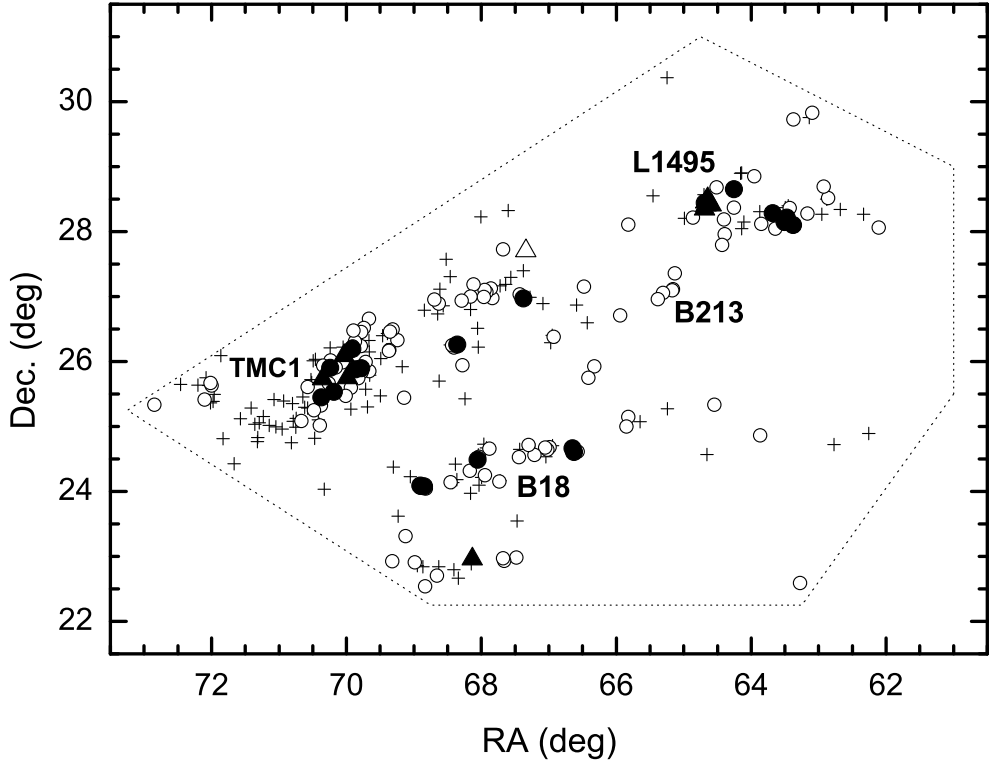}
\caption{Map of the distribution of field stars identified in this work (Table~1; J2000 coordinates). Dotted lines denote the boundaries of the survey area. Symbols used for field stars are coded by visual extinction: plus signs ($2<\av<5$); open circles ($5<\av<10$); filled circles ($\av>10$). The YSO candidate J04292083+2742074 is denoted by an open triangle. Ten highly reddened YSOs listed in Table~2 are also plotted (filled triangles; six of them are clustered together in L\,1495). The vertices of the hexagonal survey area are located at RA,~Dec.~(2000, counterclockwise from the left):  
04h~53m, $+25^\circ\ 15'$;
04h~35m, $+22^\circ\ 15'$;
04h~13m, $+22^\circ\ 15'$;
04h~04m, $+25^\circ\ 30'$;
04h~04m, $+29^\circ\ 00'$; and
04h~19m, $+31^\circ\ 00'$.
\label{fig1}}
\end{figure}

\clearpage
\begin{figure}
\centering
\includegraphics[width=10cm, angle=0]{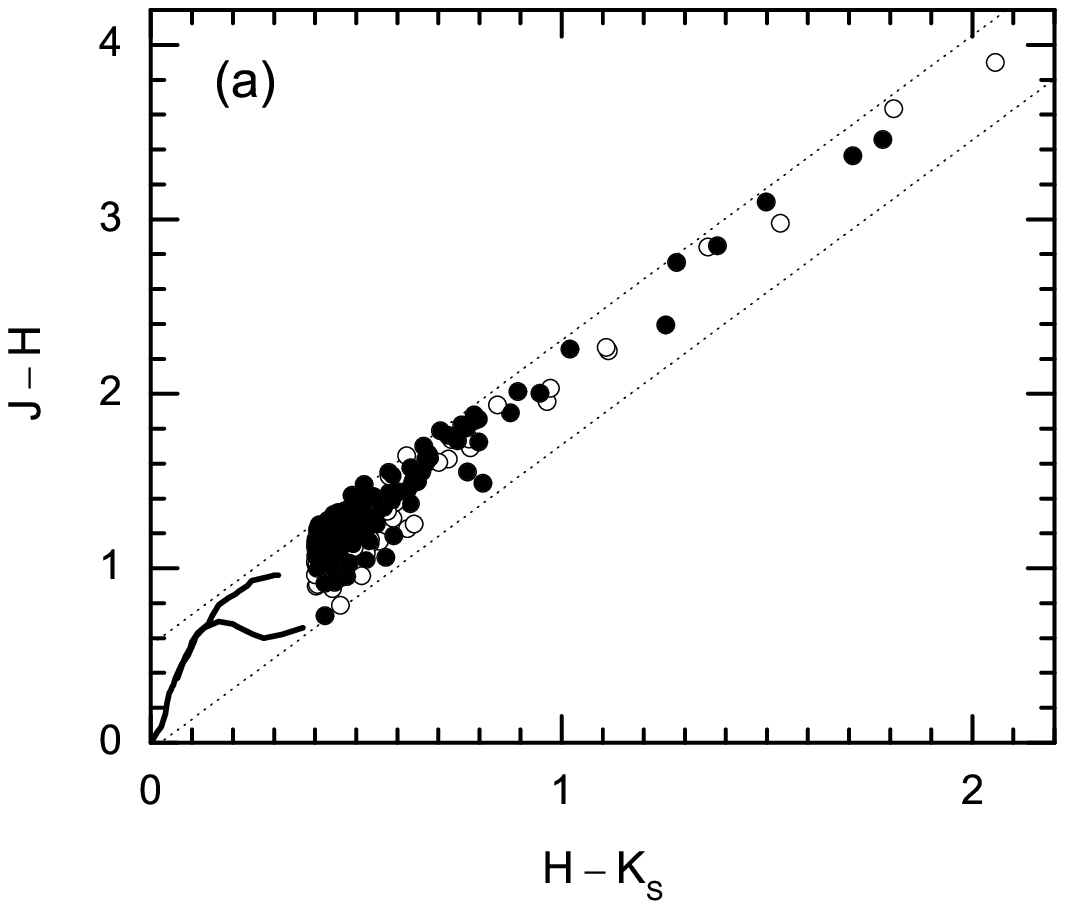}
\includegraphics[width=10cm, angle=0]{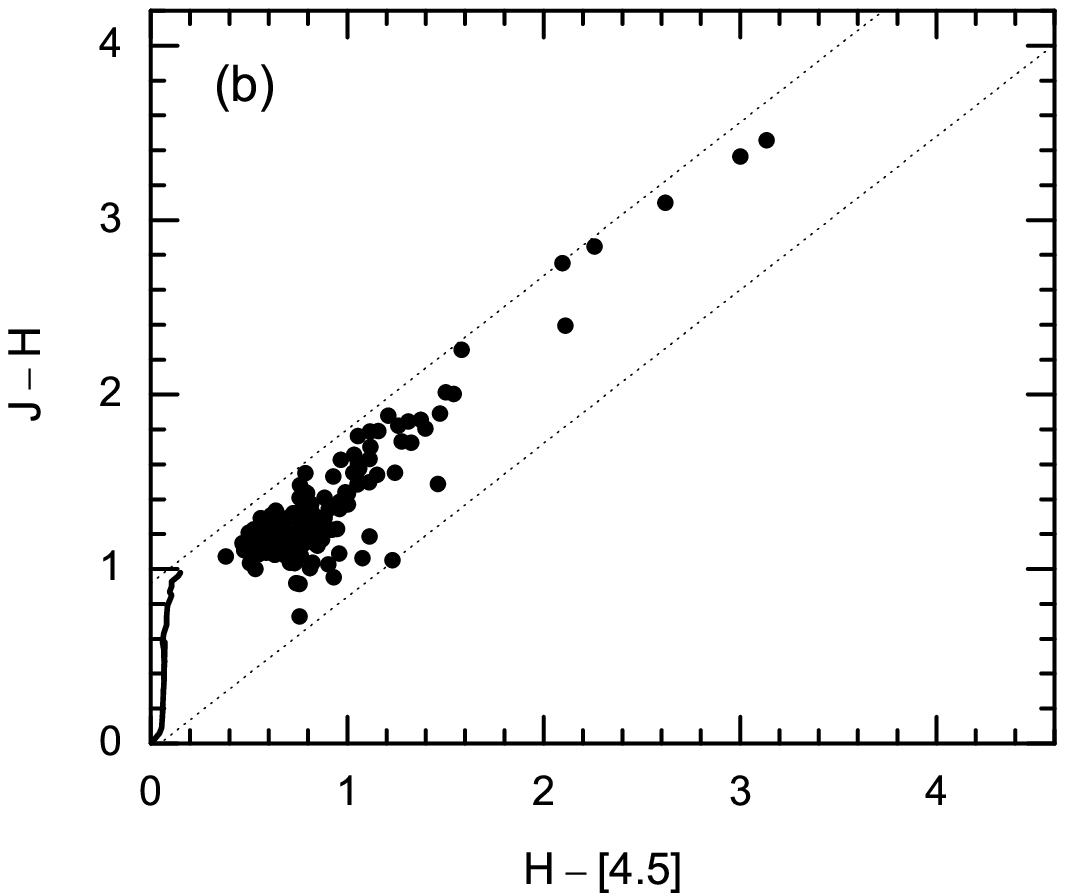}
\caption{Color-color diagrams for candidate reddened field stars listed in Table~1: (a)~$J-H$ vs.\ $H-K_{\rm s}$, and (b)~$J-H$ vs.\ $H-[4.5]$. Open circles in frame~(a) denote stars that lack 4.5\mic\ photometry and therefore do not appear in frame~(b). Solid curves near the origin represent intrinsic colors for normal stars. The dotted diagonal lines in each frame are parallel to the appropriate reddening vector and indicate the approximate upper and lower boundaries of the zone occupied by normal reddened stars lacking circumstellar infrared emission.
\label{fig2}}
\end{figure}

\clearpage

\begin{figure}
\centering
\includegraphics[width=10cm, angle=0]{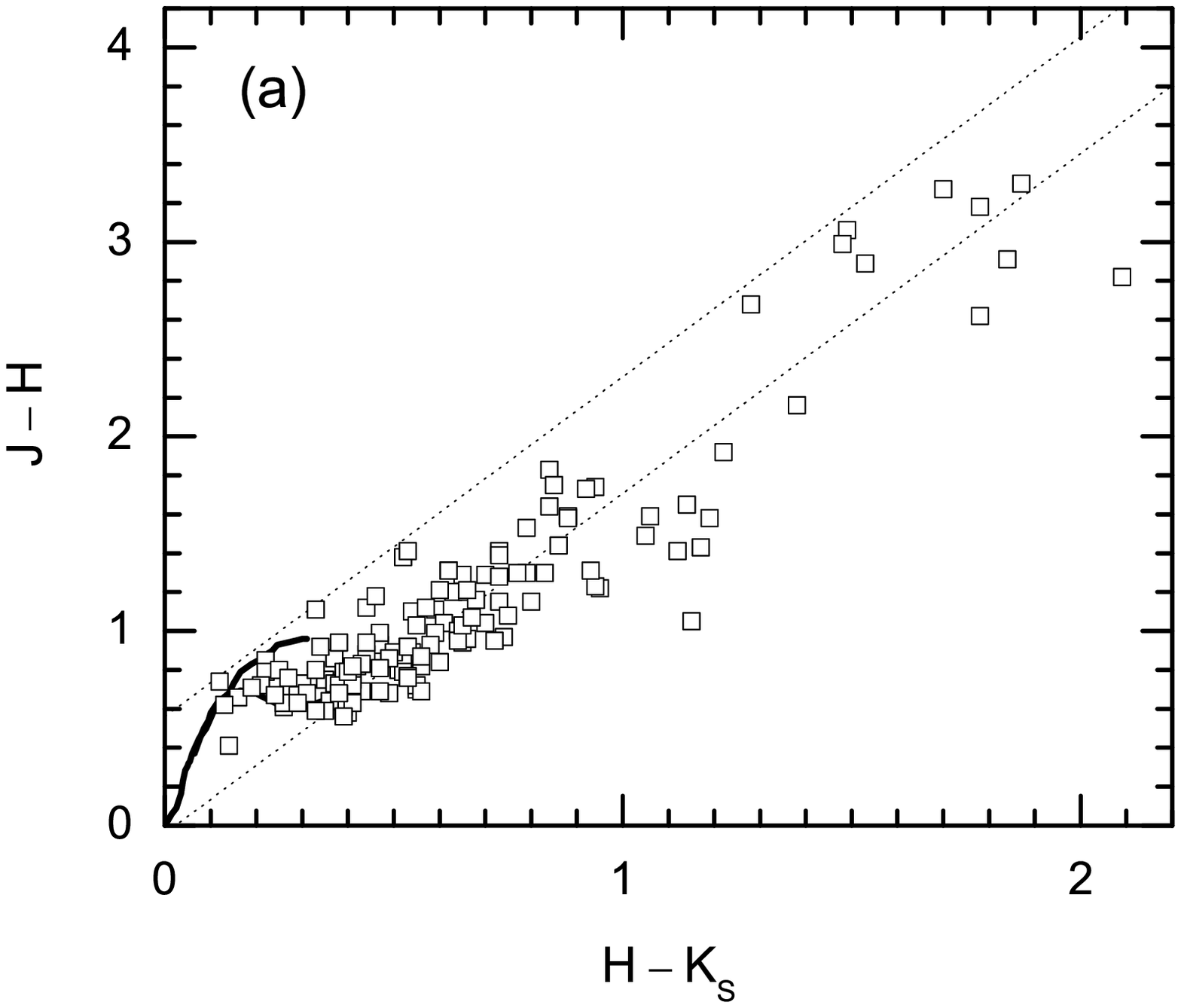}
\includegraphics[width=10cm, angle=0]{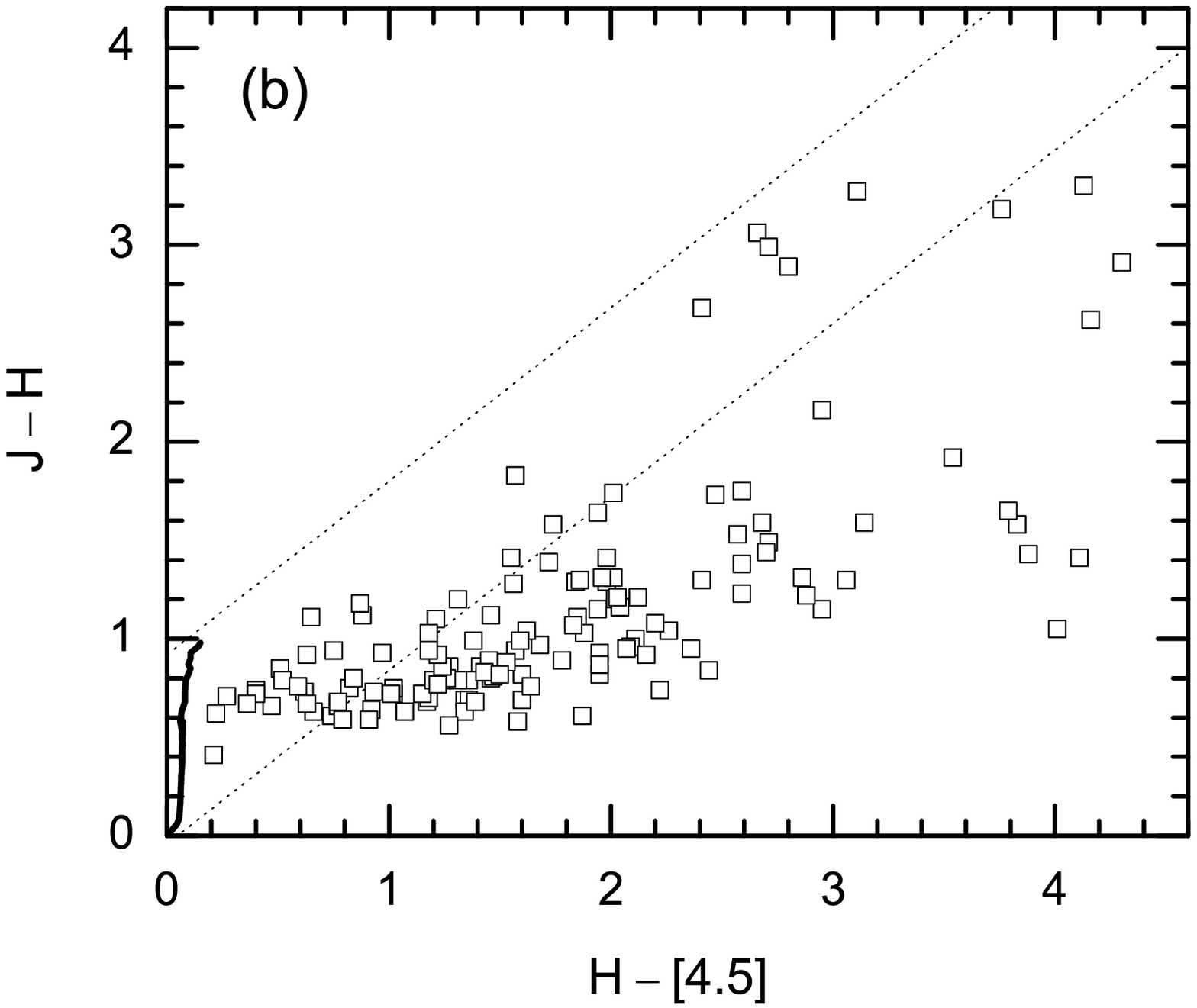}
\caption{Color-color diagrams exactly analogous to those in \fig 2, but here plotted for known members of the Taurus star-forming region. The data are from Tables~2 and 4 of Luhman \etal\ (2006); only stars with photometry available in all four relevant passbands are included. \label{fig3}}
\end{figure}

\clearpage
\begin{figure}
\centering
\includegraphics[width=12cm, angle=0]{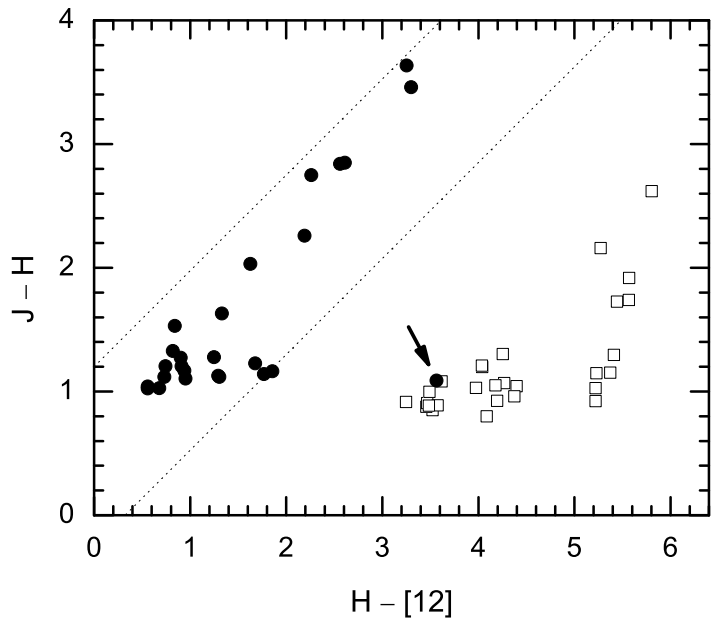}
\caption{The $J-H$ vs.\ $H-[12]$ color-color diagram. Filled circles and open squares denote candidate field stars (Table~1) and previously known YSOs (Luhman \etal\ 2006), respectively. All 12\mic\ data for YSOs are taken from the IRAS Point Source Catalog. As before, the dotted diagonal lines indicate the approximate upper and lower boundaries of the expected distribution for normal reddened stars. The datum for the anomalous object IRAS~04262+2735 (J04292083+2742074) is arrowed. \label{fig4}}
\end{figure}

\clearpage
\begin{figure}
\centering
\includegraphics[width=13cm, angle=0]{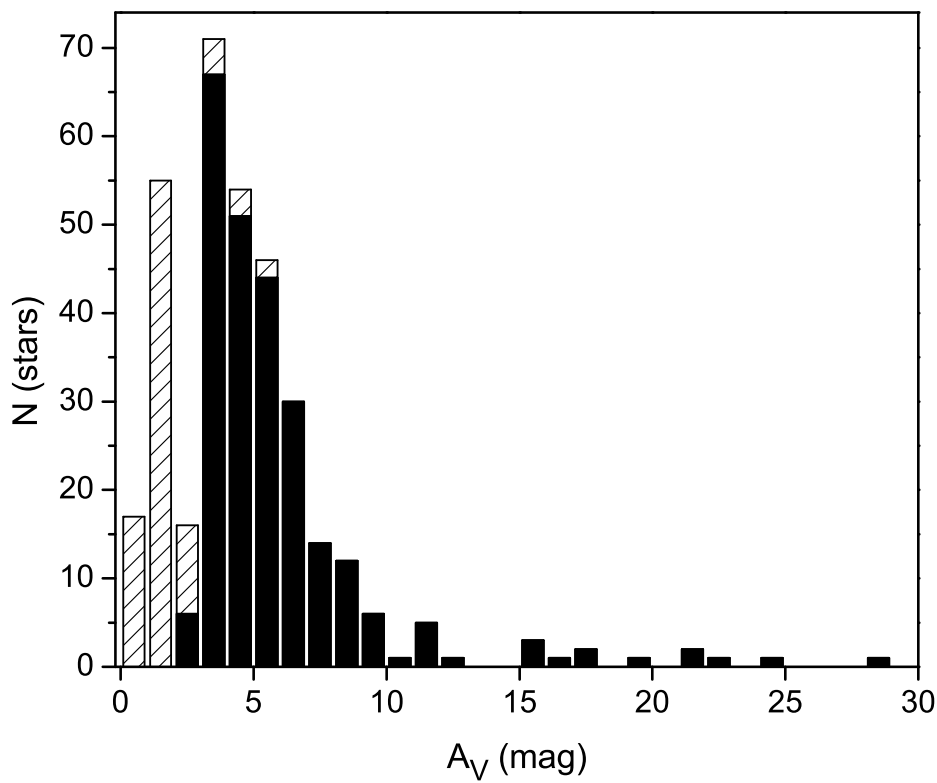}
\caption{Histogram of 249 extinction values from Table~1 divided into 1--magnitude bins (black columns). Also shown (shaded) is the effect of adding extinction data for 81 optically-selected reddened stars from the literature (Straizys \& Meistas 1980; Whittet \etal\ 2001) to our data. The total height of each column in the region of overlap (2--6~mag) sums the contributions of optical and infrared samples. \label{fig5}}
\end{figure}

\clearpage
\begin{figure}
\centering
\includegraphics[width=15cm, angle=0]{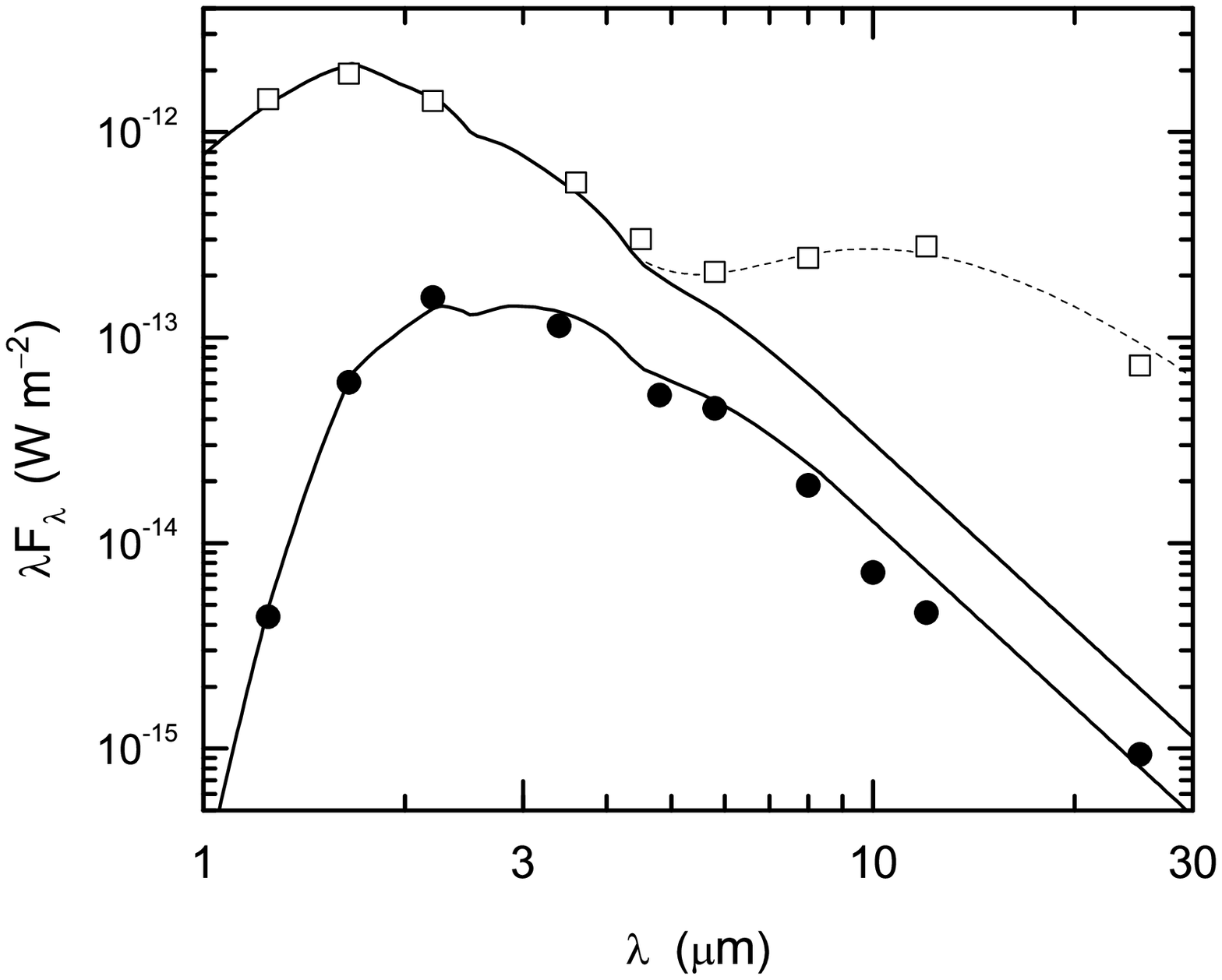}
\caption{Spectral energy distribution for the new YSO candidate IRAS~04262+2735 (J04292083+2742074, open squares) compared with that of the highly reddened field star Elias~16 (J04393890+2611250, solid circles), combining 2MASS, IRAC and IRAS photometry. Additional ground-based 3.4--10\mic\ photometry from Elias (1978) is included for the latter object as three of the four IRAC bands are saturated. Photometric errors are comparable with the size of the plotting symbols. A color correction appropriate to a 4500~K blackbody has been applied to the IRAS data for each source. Flux data for Elias~16 have been divided by a factor of 50 for display. Models (curves) were constructed assuming each star to have an intrinsic spectrum with $T_{\rm eff} = 4500$~K (Kurucz 1992) subject to extinction $\av = 5$~mag (IRAS~04262+2735) and $\av = 24$~mag (Elias~16). The dashed curve shows the effect of adding thermal emission from warm dust ($T_{\rm dust}=350$~K) to the model for IRAS~04262+2735, scaled to match the observed mid-infrared flux. \label{fig6}}
\end{figure}

\clearpage

\clearpage
\begin{thebibliography}{}
\bibitem[]{be1} Bessell, M.S. \& Brett, J.M. 1988, \pasp, 100, 1134
\bibitem[]{bo0} Boogert, A.C.A., et al. 2004, \apjs, 154, 359
\bibitem[]{bo1} Bowey, J.E., Adamson, A.J., \& Whittet, D.C.B. 1998, \mnras, 298, 131
\bibitem[]{ca0} Cambr\'esy, L. 1999, \aap, 345, 965
\bibitem[]{ca1} Cardelli, J.A., Clayton G.C., \& Mathis J.S. 1989, \apj, 345, 245
\bibitem[]{ca2} Carpenter, J.M. 2001, \aj, 121, 2851
\bibitem[]{di1} Dickens, J.E., Langer, W.D., \& Velusamy, T. 2001, \apj, 558, 693
\bibitem[]{el1} Elias, J.H. 1978, \apj, 224, 857
\bibitem[]{fa1} Fazio, G.G., et al. 2004, \apjs, 154, 10
\bibitem[]{go0} Gomez, M., Kenyon, S.J., \& Hartmann, L. 1994, \aj, 107, 1850
\bibitem[]{go1} Goodman, A.A., Jones, T.J., Lada, E.A., \& Myers P.C. 1992, \apj, 399, 108
\bibitem[]{gu1} G\"udel, M., et al. 2007, \aap, 468, 353
\bibitem[]{gu2} Gutermuth, R.A., et al. 2004, \apjs, 154, 374
\bibitem[]{ho1} Hough, J.H., et al. 1988, \mnras, 230, 107
\bibitem[]{ir1} IRAS Explanatory Supplement 1988, NASA Publication RP--1190, ed. C.A.~Beichman et al. 
				(NASA, Washington DC)
\bibitem[]{it1} Itoh, Y., Tamura, M., \& Gatley, I. 1996, \apjl, 465, L129
\bibitem[]{jh1} Jones, B.F., \& Herbig, G.H. 1979, \aj, 84, 1872
\bibitem[]{ke1} Kenyon, S.J., Dobrzycka, D., \& Hartmann, L. 1994, \aj, 108, 1872
\bibitem[]{ku1} Kurucz, R.L. 1992, in IAU Symp.\ 149, The Stellar Populations of Galaxies, eds.\  
			    B.~Barbuy \& A.~Renzini (Dordrecht: Kluwer), 225
\bibitem[]{lo1} Loinard, L., Mioduszewski, A.J., Rodriguez, L.F., Gonz\'alez, R.A., Rodriguez, M.I.,
				\& Torres, R.M. 2005, \apjl, 619, L179
\bibitem[]{lu1} Luhman, K.L., Whitney, B.A., Meade, M.R., Babler, B.L., Indebetouw, R., Bracker, S., 
 				\& Churchwell, E.B. 2006, \apj, 647, 1180
\bibitem[]{mi1} Mizuno, A., Onishi, T., Nagahama, T., Ogawa, H., \& Fukui, Y. 1995 \apjl, 445, L161
\bibitem[]{mu1} Murakawa, K., Tamura, M., \& Nagata, T. 2000, \apjs, 128, 603
\bibitem[]{pa0} Padgett, D., et al. 2007, AAS Abstract, 211, 29.04
\bibitem[]{pa1} Padoan, P., Cambr\'esy, L.,. \& Langer, W. 2002, \apj, 580, L57
\bibitem[]{pr1} Pratap, P., Dickens, J.E., Snell, R.L., Miralles, M.P., Bergin, E.A., Irvine, W.M., 
				\& Schloerb, F.P. 1997, \apj\ 486, 862
\bibitem[]{sc1} Scelsi, L., Maggio, A., Micela, G., Briggs, K., \& G\"udel, M. 2007, \aap, 473, 589
\bibitem[]{sh1} Shenoy, S.S. 2003, Ph.D.\ thesis, Rensselaer Polytechnic Institute
\bibitem[]{sk1} Skrutskie, M.F., et al. 2006, \aj, 131, 1163
\bibitem[]{so1} Sonnentrucker, P., et al. 2008, \apj, in press
\bibitem[]{st0} Strai\v{z}ys, V., \& Mei\v{s}tas, E. 1980, Acta.\ Astron., 30, 541
\bibitem[]{st1} Strom, K.M., \& Strom, S.E. 1994, \apj, 424, 237
\bibitem[]{ta1} Tamura, M., Nagata, T., Sato, S., \& Tanaka, M. 1987, \mnras, 224, 413
\bibitem[]{wh1} Whittet, D.C.B., 2003, Dust in the Galactic Environment (Institute of Physics Publishing,
				Bristol, 2nd edn.)
\bibitem[]{wh2} Whittet, D.C.B., Bode, M.F., Longmore, A.J., Admason, A.J., McFadzean, A.D., Aitken, 
				D.K., \& Roche, P.F. 1988, \mnras, 233, 321
\bibitem[]{wh3} Whittet, D.C.B., Gerakines, P.A., Hough, J.H., \& Shenoy, S.S. 2001, \apj, 547, 872
\bibitem[]{wh4} Whittet, D.C.B., et al. 2007, \apj, 655, 332
\bibitem[]{wh5} Whittet, D.C.B., Hough, J.H., Lazarian, A., \& Hoang, T. 2008, \apj, in press

\end{thebibliography}
\end{document}